\documentclass{PoS}
\usepackage{amsmath}
\pdfoutput=1

\newcommand{\la}{\langle}
\newcommand{\ra}{\rangle}
\DeclareMathOperator{\diag}{diag}
\DeclareMathOperator{\Tr}{tr}
\DeclareMathOperator{\re}{Re}
\DeclareMathOperator{\im}{Im}

\title{Large-\protect$\boldmath{N}$ phase transitions in the spectrum of
  products of complex matrices }

\ShortTitle{Large-$N$ phase transitions in the spectrum of products of
  complex matrices }

\author{\speaker{Robert Lohmayer}$^a$, Herbert Neuberger$^b$ and Tilo
  Wettig$^a$ \\ 
  \llap{$^a$}Institute for Theoretical Physics, University of
  Regensburg, 93040 Regensburg, Germany\\ 
  \llap{$^b$}Department of Physics and Astronomy, Rutgers University, 
  Piscataway, NJ 08855, USA\\
  Email: \email{robert.lohmayer@physik.uni-regensburg.de},
  \email{neuberg@physics.rutgers.edu},
  \email{tilo.wettig@physik.uni-regensburg.de}} 

\abstract{It is shown that the simplest multiplicative random complex
  matrix model generalizes the large-$N$ phase structure found in the
  unitary case: A perturbative regime is joined to a nonperturbative
  regime at a point of nonanalyticity.}

\FullConference{The XXVI International Symposium on Lattice Field
  Theory \\ 
  July 14 - 19, 2008\\
  Williamsburg, Virginia, USA}

\begin{document}

\section{Introduction}

Recent numerical work provides evidence
that Wilson loops in $SU(N)$ gauge theory in two, three and four
dimensions exhibit an infinite-$N$ phase transition: Dilated from a
small size to a large one, the eigenvalue distribution of the untraced
Wilson loop unitary matrix expands from a small arc to the entire unit
circle \cite{ourjhep, three-d}.  This transition, which was discovered by Durhuus and Olesen \cite{duol}, is in the universality class of a random multiplicative ensemble of unitary matrices \cite{janik}.  In the following, we will relax the unitarity
constraint and focus on a multiplicative random complex matrix
model introduced and solved in \cite{cmplxrmt}.

\section{Basic random complex matrix model}

We define a sequence of $n$ independent and identically distributed
$N\times N$ matrices
\begin{equation}
  M_j=\exp\left(\epsilon C_j\right)\:,\quad j=1,\ldots,n
\end{equation}
with normalized Gaussian probability distribution
\begin{equation}
  P(C)d\mu(C)=e^{-N\Tr C^\dagger C} \prod_{1\leq i,j\leq N}\frac N\pi
  \underbrace{d\mu(C_{ij})}_{ d\re(C_{ij})\:d\im(C_{ij})} 
\end{equation}
which is invariant individually under $C_j\to C^*_j, -C_j$ and $C_j\to
U_j^\dagger C_j U_j,~U_j\in U(N)$.  We do not
restrict the trace of $C_j$ here since it turns out that requiring
$\det M_j=1$ has no effect on the saddle-point analysis in the large-$N$ limit.  We are interested in the distribution of the product
\begin{equation}
  W_n =\prod_{j=1}^n M_j
\end{equation}
in the limit $n\to\infty,~\epsilon\to 0$ with the parameter
$t=\epsilon^2 n$ held fixed. $t>0$ can be interpreted as a diffusion time. This model is almost identical to that
of~\cite{cmplxrmt}.

To derive a closed formula for the entire distribution of $W$ for
general $N$ is difficult; however, similarly to the unitary model,
partial information about the distribution of eigenvalues can be
obtained from the averages of characteristic polynomials.  In the
following, averages over all $C_j$ are denoted by $\la\dots\ra$. 
$\la \det(z-W_n)\ra$ carries no information, since
\begin{equation}
  \la\det(z-W_n)\ra=(z-1)^N\:.
\end{equation}
The first non-trivial case is
\begin{equation}
  Q(z,z^*)=\la |\det(z-W_n )|^2 \ra\:.
\end{equation}
Applying large-$N$ factorization to $Q$ (i.e., assuming that the average of the product can be replaced by the product of the averages) would result in holomorphic factorization,
\begin{equation}
Q(z,z^*)\rightarrow\la\det(z-W_n)\ra\la\det(z-W_n)^*\ra=|z-1|^{2N}\:,
\end{equation}
and all eigenvalues seem to have to be unity. This factorization is expected to hold only in two
distinct regions: inside a circle around zero with radius $\rho<1$ and outside a circle of radius $\rho^{-1}>1$. Therefore, the surface eigenvalue density is restricted to the annulus $\rho<|z|<\rho^{-1}$. The aim of the following is to calculate $Q$ as a function of
$t$. We shall find that the domain of eigenvalues becomes multiply connected at a critical $t=t_c$, in agreement with \cite{cmplxrmt}.

\section{Saddle-point analysis}\label{sec_saddle}

The first step in the procedure is to disentangle the non-abelian
product of matrices defining $W_n$. To this end, we introduce $2nN$
pairs of Grassmann variables
$\left\{\bar\psi_j,\psi_j,\bar\chi_j,\chi_j\right\}_{j=1,\ldots,n}$
and find
\begin{equation}
  |\det(z-W_n)|^2=\int\prod_{j=1}^n [d\bar\psi_j d\psi_j d\bar\chi_j
  d\chi_j] e^{\sum_{j=1}^n (e^\sigma \bar\psi_j\psi_j + e^{\sigma^*}
    \bar\chi_j \chi_j) } e^{-\sum_{j=1}^n (\bar\psi_j M_j\psi_{j+1} +
    \bar\chi_j M^\dagger_j\chi_{j-1})} 
\end{equation}
with $z=e^{n\sigma}$ and the convention that $\psi_{n+1}\equiv
\psi_1$, etc.

Now, the integrals over the matrices $C_j$ factorize and can be done
explicitly to sufficient accuracy in $\epsilon$.
The following equalities ought to be understood in the sense that they hold up to terms which vanish as $n\to\infty$, $\epsilon\to0$ at $t=\epsilon^2n$ fixed.
An expansion of $M=\exp(\epsilon C)$ to linear
order in $\epsilon$ is sufficient because the next term does not
contribute,
\begin{equation}
  \la e^{-\bar\psi M\psi^\prime-\bar\chi M^\dagger \chi^\prime }\ra =
  e^{-\bar\psi\psi^\prime -\bar\chi\chi^\prime -\frac{\epsilon^2}{N}
    \bar\psi\chi^\prime\bar\chi\psi^\prime}\:. 
\end{equation}

Introducing scalar complex bosonic multipliers $\zeta_j$, $j=1,\ldots,n$
allows for a separation of the quartic Grassmann terms into bilinears,
\begin{equation}
  e^{ -\frac{\epsilon^2}{N} \bar\psi\chi^\prime\bar\chi\psi^\prime}=\frac{N}{\pi}
  \int d\mu(\zeta)e^{-N|\zeta|^2} e^{-\epsilon ( \zeta \bar\psi
    \chi^\prime - \zeta^*\bar\chi \psi^\prime)}\:, 
\end{equation}
where the integration measure is $d\mu(\zeta )= d\re\zeta\:
d\im\zeta$. After shifting some indices of Grassmann variables, this
leads to
\begin{align}
  Q(z,z^\ast)&=\left(\frac{N}{\pi}\right)^{n}\int\prod_{j=1}^n [d\bar\psi_j d\psi_j d\bar\chi_j
  d\chi_j d\mu(\zeta_j)]e^{-N\sum_{j=1}^n|\zeta_j|^2}
  e^{-\sum_{j=1}^n(\bar\psi_j \psi_{j} +\bar\chi_j \chi_{j}) }\nonumber\\
  &\quad\times e^{\sum_{j=1}^n (e^\sigma
    \bar\psi_j\psi_{j-1} + e^{\sigma^*}  \bar\chi_j \chi_{j+1}) }
  e^{-\epsilon\sum_{j=1}^n (\zeta_j\bar\psi_j \chi_{j} - \zeta^*_j
    \bar\chi_{j}\psi_{j})}\:. 
\end{align}
As a result, averages over complex
matrices $C_j$ are replaced by averages over complex numbers
$\zeta_j$.  Carrying out the integrals over the Grassmann variables
makes the dependence on $N$ explicit, and we are left with
\begin{equation}
  Q(z,z^\ast)=\left(\frac{N}{\pi}\right)^{n}\int\prod_{j=1}^n [d\mu(\zeta_j)]
  e^{-N\sum_{j=1}^n|\zeta_j|^2}{\det}^N
  \begin{pmatrix}
    A & B \\ C & D  
  \end{pmatrix}\:,
\end{equation}
where
\begin{equation}
  A=e^\sigma T^\dagger -1\:,\quad D=A^\dagger\:,\quad B=-\epsilon
  Z\:,\quad C=\epsilon Z^\dagger  
\end{equation}
with $n\times n$ matrices
\begin{equation}
  T=
  \begin{pmatrix}
    0 &1 & 0 & \cdots & 0 & 0\\
    0 &0 & 1  & \cdots & 0 & 0\\
    \vdots & \vdots & \vdots &\cdots & \vdots & \vdots \\
    0 &0 & 0  & \cdots & 0 & 1\\
    1 & 0 & 0 & \cdots & 0 & 0  
  \end{pmatrix}
  \quad\text{and}\quad 
  Z=\diag(\zeta_1,\zeta_2,\ldots,\zeta_n)\:.
\end{equation}

The large-$N$ limit leads to saddle-point equations trivially
satisfied at $\zeta_j=0$ for all $j=1,\ldots,n$ since the
$\zeta_j,\zeta_j^\ast$ enter only bilinearly in
\begin{equation}
  \det\begin{pmatrix}
    A & B \\ C & D  
  \end{pmatrix}
  =|\det A|^2\det ( 1+ \epsilon^2 Z^\dagger A^{-1} Z (A^\dagger)^{-1} )\:.
\end{equation}
Where this saddle dominates we obtain
\begin{equation}
  Q(z,z^\ast)=|z-1|^{2N}
\end{equation}
and $W_n$ could be replaced by a unit matrix, which means that there
are no eigenvalues at any $z\neq1$ in the complex plane.

Comparison with numerical simulation shows that the trivial saddle point is always dominating whenever it is locally stable. At the
boundary of the local domain of stability one has a transition to
regions with non-zero surface eigenvalue density. To determine this
boundary we need to expand the integrand
around $\zeta_j=0$,
\begin{equation}
  \det\begin{pmatrix}
    A & B \\ C & D  
  \end{pmatrix}
  =|\det(1-e^\sigma T^\dagger)|^2 e^{\epsilon^2 \sum_{j,l=1}^{n}
    \zeta_j^{\ast}K_{jl}\zeta_l} 
\end{equation}
with $K_{jl}=|(A^{-1})_{jl}|^2$. The matrix $T$ implements cyclical
one-step shifts: $T^n=1$. Hence,
\begin{equation}
  A^{-1}=\frac{1}{e^\sigma T^\dagger-1} =\frac{1}{1-e^{-n\sigma}}
  \sum_{s=1}^{n} e^{-s\sigma} T^s\:, 
\end{equation}
implying that $K$ is a circulant matrix. Its eigenvalues are found to
be
\begin{equation}
  \lambda_{k}=\left |\frac{1}{1-e^{-n\sigma}}\right |^2 
  \sum_{j=1}^n e^{-j(\sigma+\sigma^*)} e^{\frac{2\pi i}{n} kj}=
  \left |\frac{1}{1-e^{-n\sigma}}\right |^2
  \frac{1-e^{-n(\sigma+\sigma^*)}}{1-e^{\frac{2\pi\imath}{n} k}
    e^{-(\sigma+\sigma^*)}} e^{-(\sigma+\sigma^*)} 
  e^{\frac{2\pi i}{n}k}\:. 
\end{equation}
The condition of local stability,
\begin{equation}
  \re(-1+\epsilon^2\lambda_{k})<0\:,\quad k=1,\ldots,n
\end{equation}
is strongest for $k=n$, and consequently the region of local stability of the
trivial saddle point is
\begin{equation}
  \epsilon^2 \frac{1}{|z-1|^2} \left (
    \frac{|z|^2-1}{|z|^{\frac{2}{n}} -1} \right ) < 1\:. 
\label{finiteStab}
\end{equation} 
Taking the limit ($n\to\infty$, $\epsilon\to0$ with $t=\epsilon^2n$ kept fixed) gives
\begin{equation}
  1 > \frac{t}{2|z-1|^2} \frac{|z|^2-1}{\log |z|}
  \label{basicStab}
\end{equation}
in agreement with \cite{cmplxrmt}. 

In contrast to (\ref{finiteStab}), the last inequality is invariant under $z\to z^{-1}$. The transition occurs when the inversion invariant point $z=-1$ on the unit circle first enters the domain of eigenvalues.
The condition for vanishing eigenvalue density at a point $z$ on the unit circle reads
\begin{equation}
  t<|z-1|^2\quad\text{for}\quad |z|=1\:.
\end{equation}
With $z=e^{i\psi}$, this is equivalent to
\begin{equation}
\cos\psi<1-\frac t2\:.
\label{unitStab}
\end{equation} 
When $t>4$  this condition is clearly violated for any
$\psi$. Consequently, the region of non-vanishing eigenvalue density
contains the whole unit circle.  On the other hand, for $t<4$ (\ref{unitStab}) will be fulfilled for some points on the unit circle. We see that the unit circle contains an arc, centered at $z=-1$ with endpoints at angles
$\psi$ satisfying $\cos(\psi)=1-\frac{t}{2}$, which lies completely in the domain of zero eigenvalue density. 
The domain of non-vanishing eigenvalue density becomes multiply connected for $t>4$. 

\section{Generalized model}

The basic random complex matrix model can be generalized to
interpolate between the cases where the individual factors are unitary
or hermitian. To this end, we write each matrix $C$ as a linear
combination
\begin{equation}
  C=H_1+ i H_2\quad \text{with}\quad H_{1,2}^\dagger=H_{1,2}
\end{equation}  
and introduce two weight factors $\omega_{1,2}>0$ in its probability
distribution,
\begin{equation}
  P(C)=\mathcal{N} e^{-N\left( \frac1{2\omega_1}\Tr
      H_1^2+\frac1{2\omega_2}\Tr H_2^2\right)}\:. 
\end{equation}
Setting $\omega_1=\omega_2=\frac12$ we get back to the basic model, and for $\omega_1\to 0$ we are in the unitary case, where the spectrum is
restricted to the unit circle.

For $\omega_1\ne\omega_2$, 
\begin{equation}
  \langle \det(z-W_n)\rangle = J(z)
\end{equation}
is no longer equal to $(z-1)^N$, but to a more complicated polynomial
in $z$. The polynomial $J(z)$ is completely determined by the
two-point function of the matrix $C$, which depends only on the
difference $\omega_2-\omega_1$. Therefore we can simply go to the
unitary case with $\omega_1=0$, for which the polynomial $J(z)$ is
known from previous work on products of unitary matrices
\cite{three-d}, and absorb the dependence on $\omega_2-\omega_1$ in
$t$ by a rescaling,
\begin{equation}
t_\pm\equiv t(\omega_2\pm\omega_1),\quad t_+ \ge |t_- |\:.
\end{equation}
Again, where holomorphic factorization,  
\begin{equation}
  Q(z,z^\ast)=\langle |\det(z-W_n )|^2 \rangle = |J(z)|^2\:,
\end{equation}
holds we have no finite surface charge density. To determine for which
values of $z$ this formula no longer gives the correct answer we apply
a strategy similar to the one presented in Sec.~\ref{sec_saddle}. The
only complication is that more quartic Grassmann terms need to be
decoupled. Thus, integrals over real multipliers
$\xi_j$, $\eta_j$, $j=1,\ldots,n$ have to be introduced in addition to the
complex noise factors $\zeta_j$. However, the complex variables again
enter only as bilinears $\zeta_j\zeta_k^\ast$, leading to a trivial
saddle point $\zeta_j=0$ at large $N$. The eigenvalue density vanishes
where this saddle dominates because the remaining $\xi$ and $\eta$
integrals factorize, resulting in holomorphic factorization for
$Q$. Therefore, only the structure of the $\zeta$ saddle determines if
$z$ is in a chargeless region, but the dominating saddle points of the
$\xi,\eta$ integrals affect the local stability of the trivial saddle
point at $\zeta_j=0$. The condition for local stability and vanishing
eigenvalue density in the desired limit ($\epsilon\to 0$, $n\to \infty$ with $t=\epsilon^2n$ kept fixed) is eventually found to be
\begin{equation}
  1 > \frac{t_+}{2|{\hat z}-1|^2} \frac{|{\hat z} |^2-1}{\log |{\hat
      z}|}\: .
\end{equation}
This inequality is similar to Eq.~\eqref{basicStab} for the basic
model, with $t$ replaced by $t_+$ and $z$ replaced by $\hat z$, where
the variable $\hat z$ is related to the original variable $z$ via
\begin{equation}
  z=\hat z e^{-t_- \hat u}\quad\text{with}\quad \hat u=
  \frac12 \left(\frac{\hat z+1}{\hat z -1}\right)\:. 
\end{equation}
The topological transition in the domain of non-vanishing eigenvalue
density thus occurs at $t_+=4$ for the generalized model.

\section{Numerical results}

\begin{center}
  \begin{figure}[ht]
    \includegraphics[height=0.33\textwidth,angle=90]{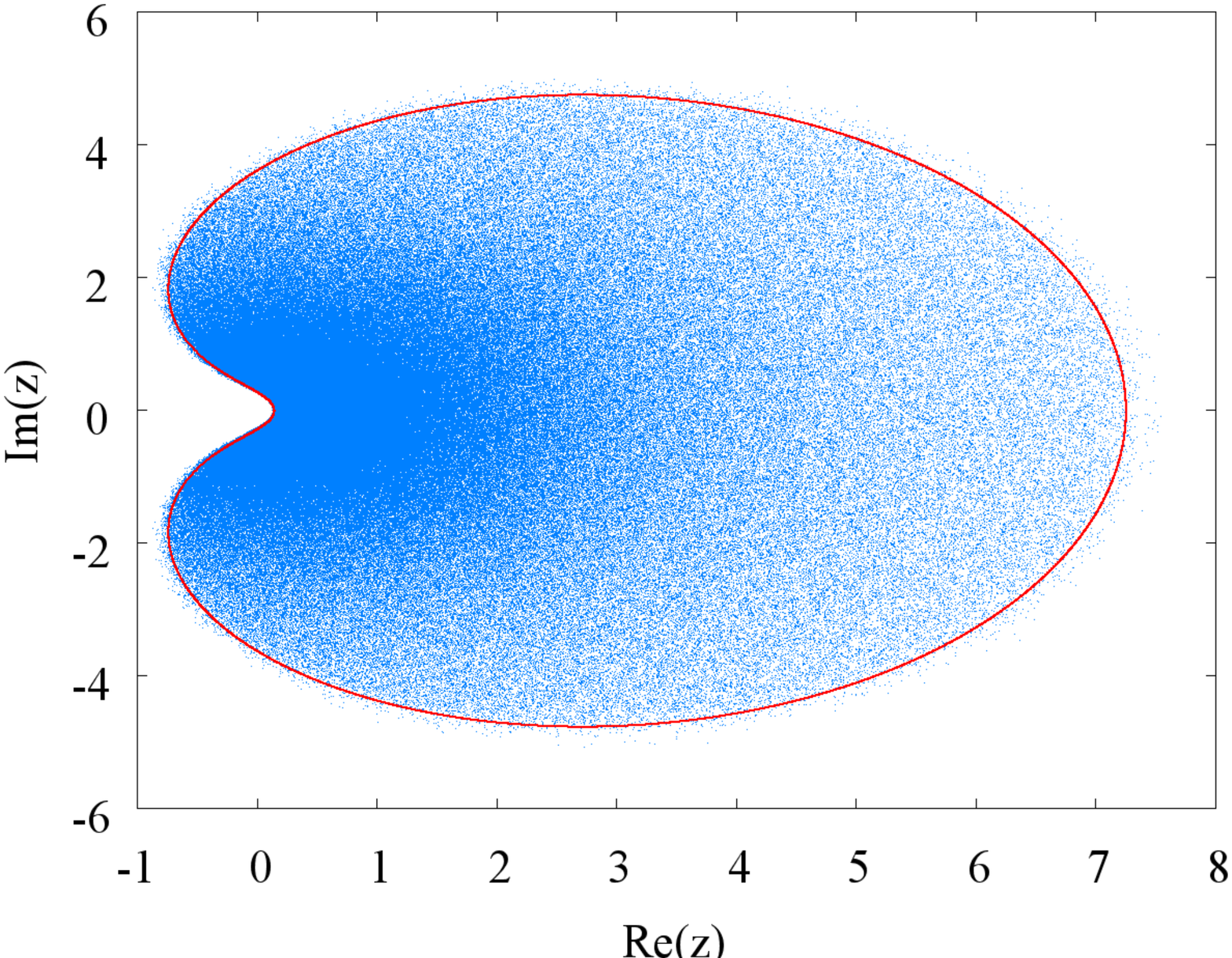}\hfill
    \includegraphics[height=0.33\textwidth,angle=90]{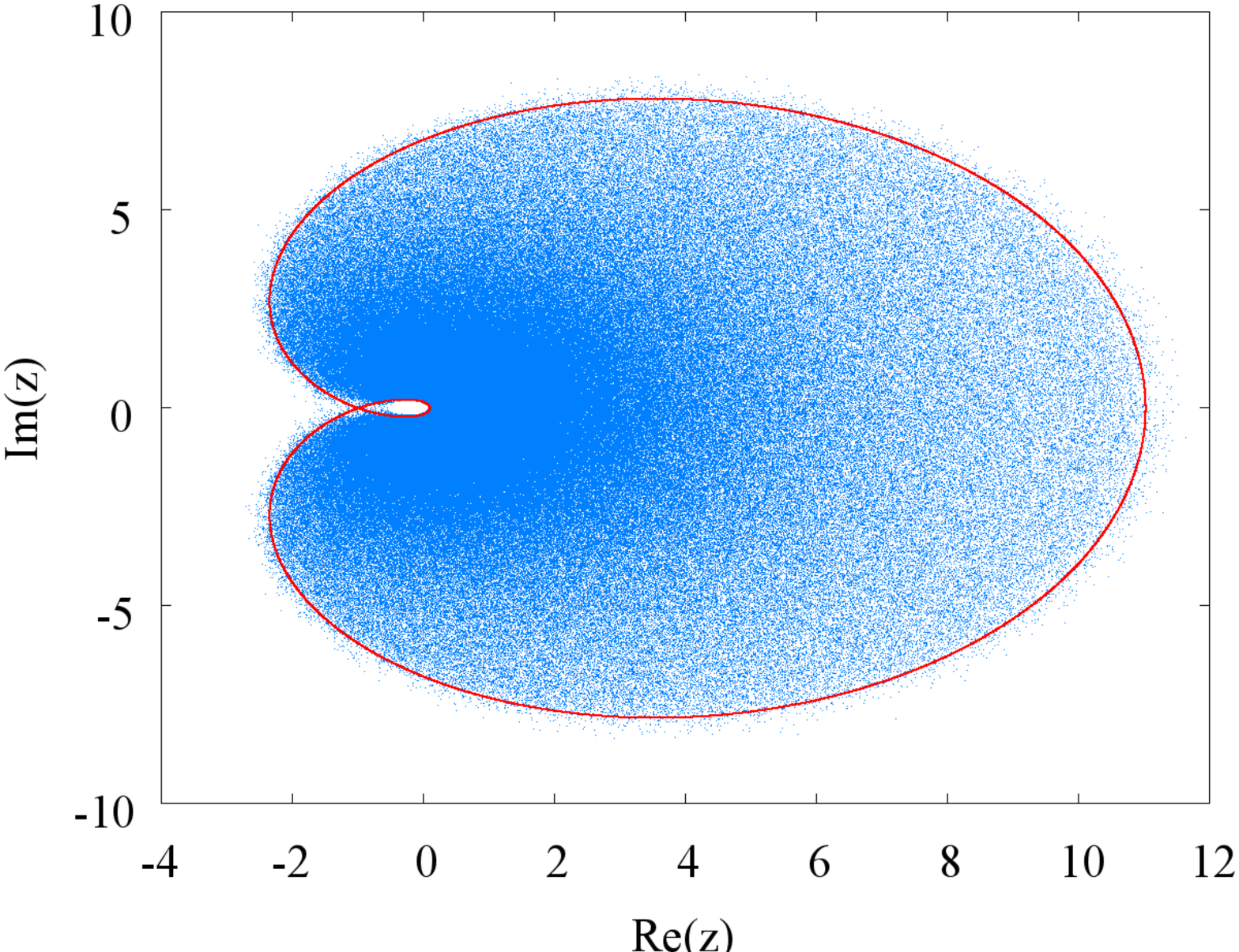}\hfill
    \includegraphics[height=0.33\textwidth,angle=90]{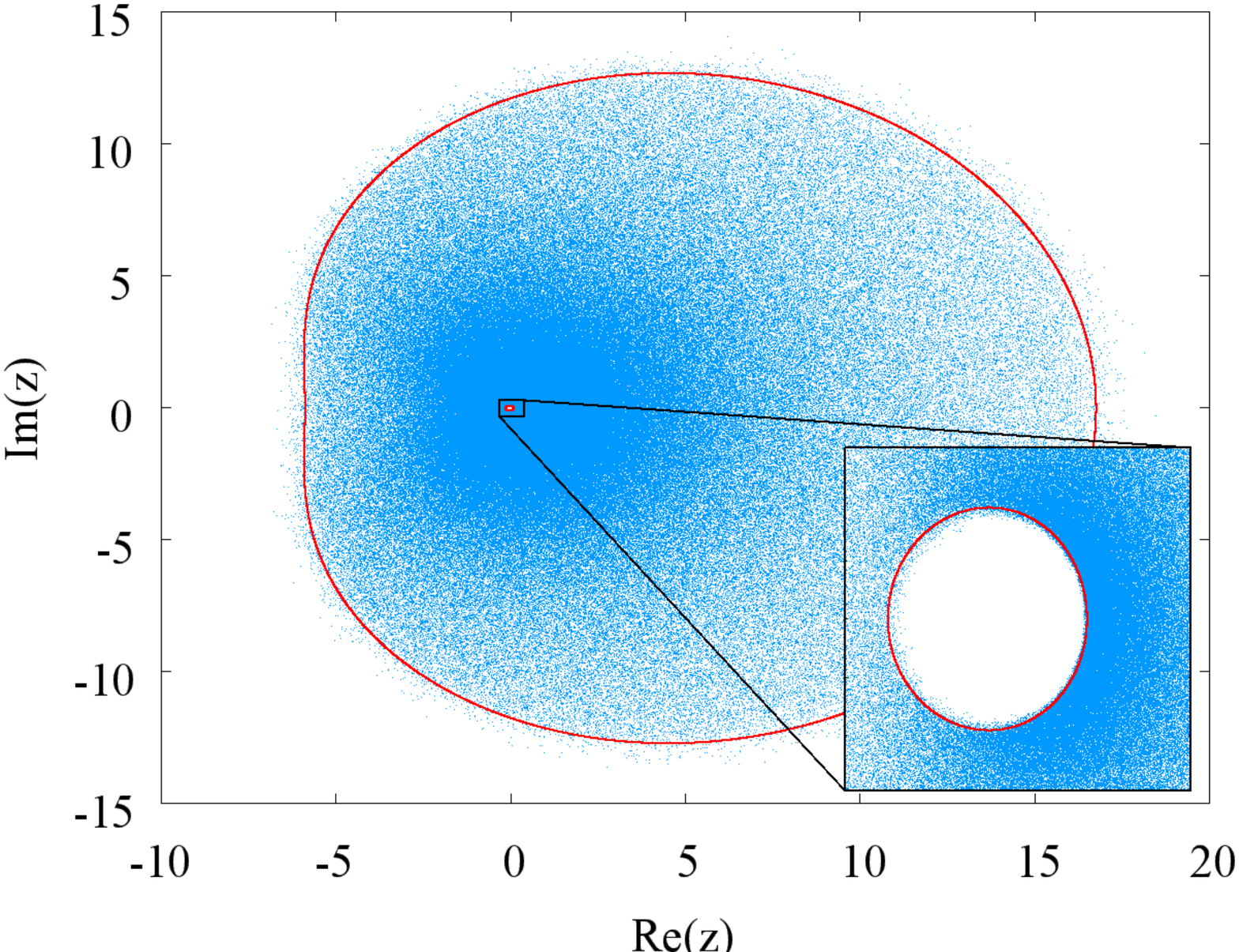}
    \caption{Scatter plot of the eigenvalues of $W_n$ for
      $\omega_1=\omega_2=\frac12$ and $t=3$ (left), $t=4$ (middle),
      $t=5$ (right).}
    \label{fig1}
  \end{figure}
\end{center}

\begin{center}
  \begin{figure}[ht]
    \includegraphics[height=0.33\textwidth,angle=90]{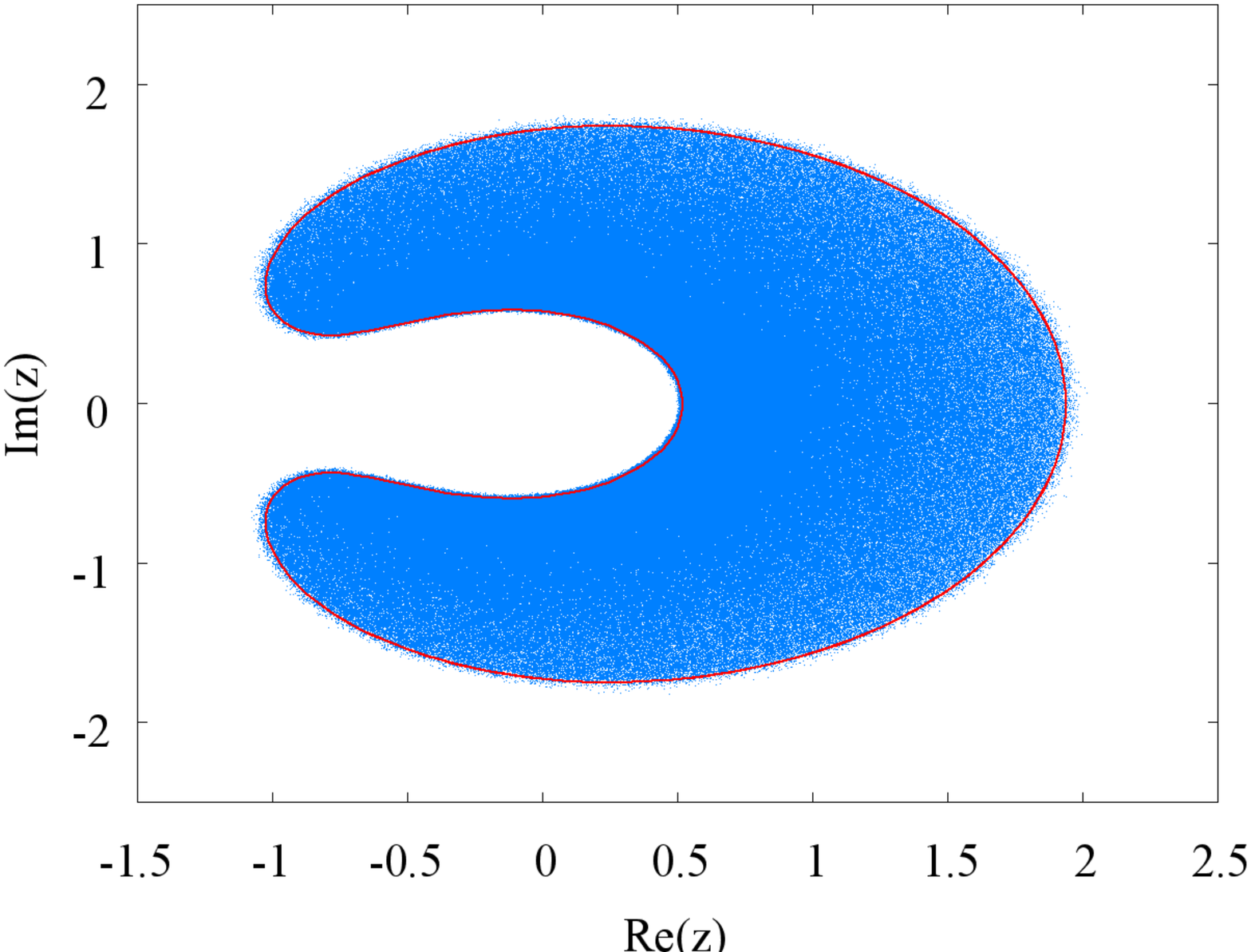}\hfill
    \includegraphics[height=0.33\textwidth,angle=90]{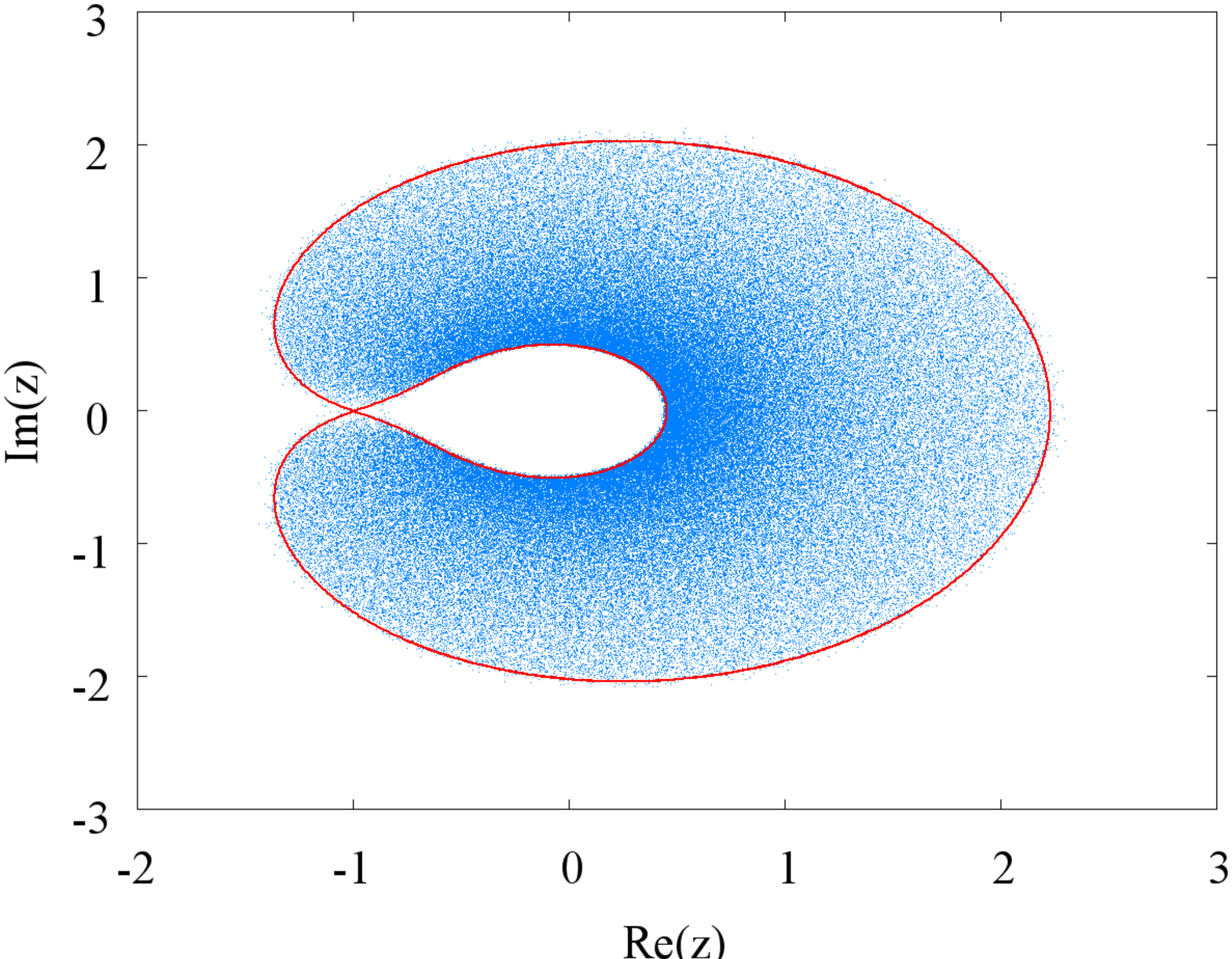}\hfill
    \includegraphics[height=0.33\textwidth,angle=90]{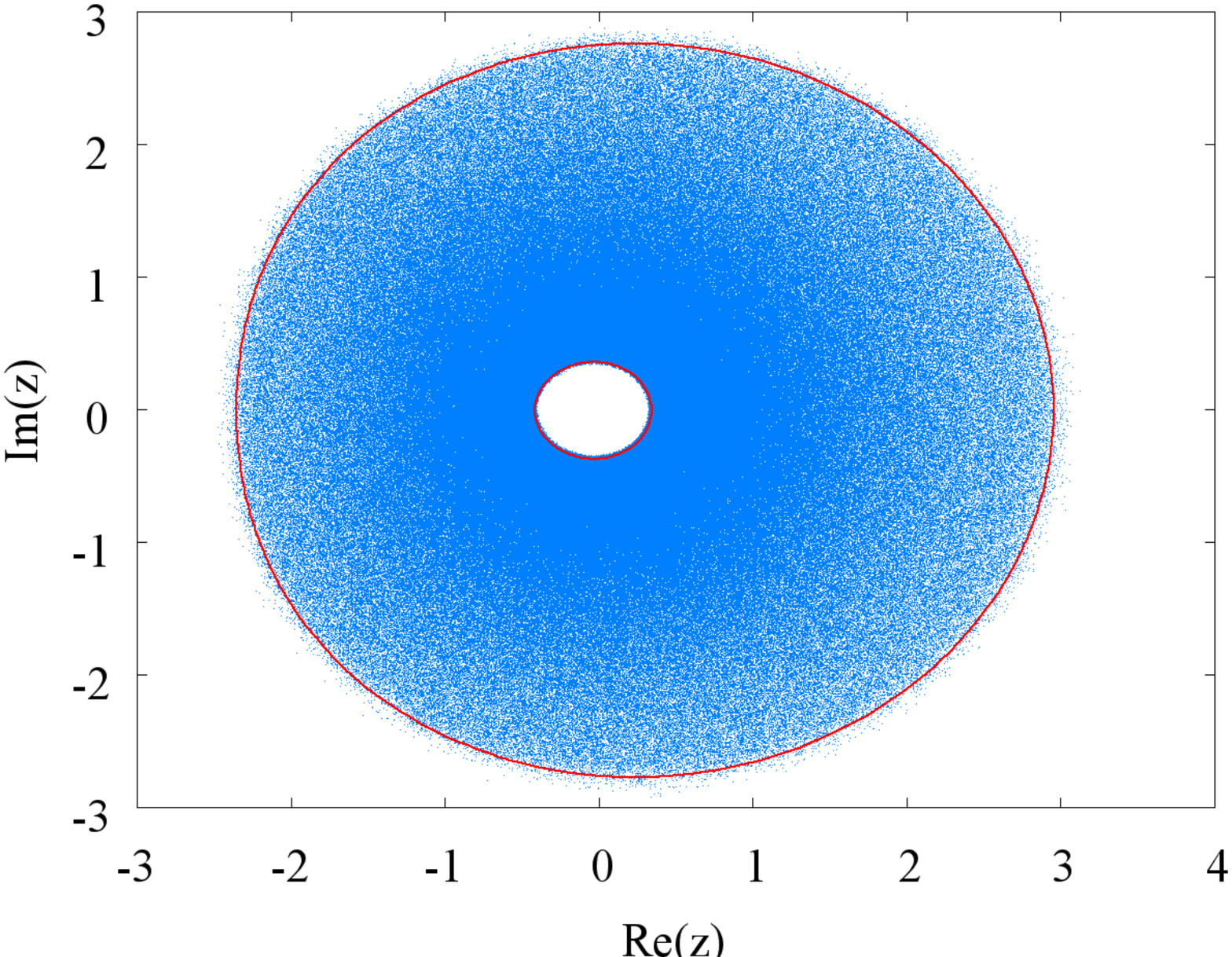}
    \caption{Same as Fig.~\protect\ref{fig1} but for
      $\omega_1=\frac1{10}$, $\omega_2=\frac12$ and $t=5$ (left),
      $t=\frac{20}3$ (middle), $t=10$ (right).}
    \label{fig2}
  \end{figure}
\end{center}

Since we did not explicitly identify the competing non-trivial saddle
points, nor determine the global stability of the trivial saddle, to establish the transition more evidence is needed, which we
provide by numerical simulations. It is more convenient numerically to  work with
the linear model $M=1+\epsilon C$ introduced in \cite{cmplxrmt},
instead of our exponential model $M=\exp(\epsilon C)$. Repeating the analysis for this model, it turns out that the boundaries of non-vanishing
eigenvalue density are equivalent up to a scaling by a factor of
$\exp(-t_-/2)$, which is equal to unity for the basic random complex
matrix model. The following figures show results of numerical
simulations for the linear model performed with matrix dimension
$N=2000$ and $n=2000$ factors in each matrix product for ensembles
consisting of about $500$ product matrices. Figure~\ref{fig1}
corresponds to the basic model. The boundaries obtained from the
stability analysis, indicated by the solid lines in red, are in very
good agreement with numerical data given by the blue points. Figure~\ref{fig2} shows results of numerical
simulations for $\omega_1=\frac 1{10},~\omega_2=\frac 12$ (data
points as well as boundaries are scaled with the corresponding factor
of $\exp(-t_-/2)$). The
topological transition occurs at $t=20/3$, which corresponds to
$t_+=\frac{20}{3}\left(\frac12+\frac1{10}\right)=4$ in agreement with
the prediction.

\section{Conclusion}

Our main objective in re-analyzing the model of~\cite{cmplxrmt} is our
conjecture that it is a universal representative of the large-$N$
phase structure of classes of complex matrix Wilson loops.  A
generalization of the probability distribution allows for an
interpolation between the cases where the individual factors are
hermitian and unitary.  We confirm, analytically and numerically, that
a topological transition in the domain of non-vanishing eigenvalue
density indeed occurs.  Like in the unitary case, we would like to
study in more detail the approach to infinite $N$ and see what the
matrix model universal features of this transition are.  Since a full
analysis keeping the exact $N$ dependence is complicated this has not
been carried to completion yet.

\section*{Acknowledgments}
We acknowledge support by BayEFG (RL), by DOE grant \#
DE-FG02-01ER41165 and SAS of Rutgers (HN), and by DFG (TW).

\end{document}